\documentclass[12pt,oneside,reqno]{amsart}

\usepackage[T1]{fontenc}
\usepackage{array}
\usepackage{caption}
  \usepackage[labelfont=md]{subcaption}
\usepackage{enumitem}
\usepackage[a4paper,tmargin=1in,bmargin=1.2in ,lmargin=1in,rmargin=1in,headheight=0in,headsep=0in,footskip=1.5\baselineskip]{geometry}
\usepackage{natbib} 
\usepackage{setspace}
\usepackage{tabularx} 
\usepackage{verbatim} 
\usepackage{mleftright}
\usepackage{paralist}
\usepackage{comment}

\usepackage[colorlinks=true,linkcolor=blue,citecolor=blue]{hyperref}

\theoremstyle{definition}
\newtheorem{theorem}{Theorem}

\newtheorem{lemma}{Lemma}
\newtheorem{corollary}{Corollary}

\newtheorem{example}{Example}
\newtheorem*{altplain}{\theoremname}

\theoremstyle{definition}
\newtheorem{definition}{Definition}

\newtheorem*{altdef}{\theoremname}

\theoremstyle{remark}
\newtheorem{claim}{Claim}

\newcommand{\theoremname}{Please Insert Name}

  {
    \renewcommand{\theoremname}{#1}
    \begin{altplain}
  }
  {\end{altplain}}

  {
    \renewcommand{\theoremname}{#1}
    \begin{altdef}
  }
  {\end{altdef}}

\newcommand{\R}{\mathbb{R}}
\newcommand{\eps}{\varepsilon}
\newcommand{\set}[1]{\left\{#1\right\}}
\newcommand{\cBr}[2]{\left(#1\mid#2\right)}

\newcommand{\citeapos}[1]{\citeauthor{#1}'s\ \citeyearpar{#1}}

\begin{document}

\title{Correlated Perfect Equilibrium}
\thanks{A version of part of this paper was submitted to the University of Queensland as Huang's thesis in partial fulfillment of the Honours degree requirement. The authors thank the editor and three anonymous referees for their helpful comments and suggestions, as well as seminar participants at the University of Bristol, Virginia Tech, the 2020 Australasian Economic Theory Workshop, and FUR 2024. Man acknowledges support from the Australian Research Council Discovery Early Career Researcher Award DE180101452.}

\author{J.~Jude Kline}
\thanks{School of Economics, University of Queensland. 
E-mail: \href{mailto:j.kline@uq.edu.au}{\texttt{j.kline@uq.edu.au}}}

\author{Wanying Huang}
\thanks{Department of Economics, Monash University. 
E-mail: \href{mailto:kate.huang@monash.edu}{\texttt{kate.huang@monash.edu}}}

\author{Priscilla Man}
\thanks{School of Economics, University of Queensland. 
E-mail: \href{mailto:t.man@uq.edu.au}{\texttt{t.man@uq.edu.au}}}

\date{August 31, 2026}

\begin{abstract}
We propose a refinement of correlated equilibrium based on common perturbations of the mediator’s recommendation distribution, which we call \emph{correlated perfect equilibrium} (CPE). In finite games, the set of CPEs is nonempty and forms a finite union of convex polyhedra. Like perfect equilibrium, a CPE never assigns positive probability to any weakly dominated strategy. We provide a dual representation of CPE, which yields an iterative procedure for identifying CPEs. These perturbations provide a common source of off-equilibrium beliefs, a feature that distinguishes CPE from two existing refinements of correlated equilibrium\textemdash acceptable correlated equilibrium (Myerson, 1986) and perfect direct correlated equilibrium (Dhillon and Mertens, 1996)\textemdash as we demonstrate through examples.\\

\noindent \textbf{Keywords:} Correlated Equilibrium, Perfection, Refinement, Normal Form Games 

\noindent \textbf{JEL classification codes:} C72
\end{abstract}

\maketitle

\newpage

\section{Introduction}
A central solution concept in game theory is correlated equilibrium \citep{Aumann1974}.%
\footnote{It can be understood either as an expression of Bayesian rationality with a common prior \citep{Aumann1987}, or as a joint rationality condition \citep{NauMcCardle1990}. See also \citet{brandl2026axioms} for a recent axiomatization of correlated equilibrium. As a generalization of Nash equilibrium, correlated equilibrium is also much easier to compute; see, e.g., \citet{papadimitriou2008computing, daskalakis2009complexity}.} 
It is often described as a coordination plan in which a mediator draws a joint strategy profile from a distribution and privately recommends to each player her component. While each player observes only her own recommendation, the distribution from which recommendations are drawn is common knowledge. Each player then conditions this distribution on her own recommendation to form beliefs about her opponents' play. A distribution is a correlated equilibrium if each player finds it optimal to follow her recommendation, assuming others do so as well. Correlated equilibrium, however, is silent about how players form beliefs following events that occur with zero probability under this distribution. This raises a natural question: should off-equilibrium beliefs also be derived from a common distribution?

In this paper, we propose a new solution concept, the \emph{correlated perfect equilibrium} (CPE), in which the distribution over recommendations is subject to perturbations. Fundamentally, these perturbations provide a \emph{common} source for off-equilibrium beliefs: players evaluate their payoffs upon ``surprises'' by conditioning on the same perturbed distribution over recommendation profiles. One can interpret these perturbations as mistakes or noise in the coordination device, with players nevertheless following the resulting recommendations. This common off-equilibrium belief distinguishes CPE from existing refinements of correlated equilibrium based on players' trembles, which may generate player-specific private information.

Formally, a CPE is a distribution over strategy profiles that is the limit of a sequence of completely mixed distributions. In such a sequence, any strategy recommended with positive probability in the limit must remain a best response throughout. This formulation is analogous to normal-form perfection for Nash equilibrium \citep{Selten1975}, in which equilibrium strategies must remain best responses throughout a sequence of independently and completely mixed distributions (see \citeapos{Myerson1978Proper} formulation). Instead, CPE allows the distributions in the perturbing sequence to be correlated, as the concept is applied to a correlated environment. 

We obtain the following results. First, as a refinement of correlated equilibrium, we establish the existence of CPE in finite normal-form games. This follows from the fact that perfect equilibria always exist in finite games \citep{Selten1975} and that every perfect equilibrium is itself a CPE (Lemma~\ref{prop:existence}). Moreover, CPE inherits a desirable property of normal-form perfection: unlike correlated equilibrium, it never assigns positive probability to weakly dominated strategies (Lemma~\ref{lem:weak}).

Our main result (Theorem~\ref{thm:dual}) provides a dual characterization of the set of CPEs. This characterization shows that this set can be expressed as a finite union of convex polyhedra and yields a tractable iterative method for identifying these equilibria. It also connects CPE to the dual characterization of correlated equilibrium, which requires a ``no arbitrage'' condition at every strategy profile in the equilibrium support \citep{NauMcCardle1990}. Our dual characterization strengthens this condition by extending the no-arbitrage requirement to strategy profiles that lie in the product of the players' individual equilibrium supports. This includes strategy profiles that receive zero probability in equilibrium.

There are three features that distinguish CPE from existing refinements of correlated equilibrium. First, because CPE places perturbations directly on the distribution over recommendation profiles, it accommodates correlation across recommendations without imposing additional assumptions on their relative likelihoods.%
\footnote{A related notion in the Nash-equilibrium setting is \emph{c-perfect equilibrium} \citep*{fudenberg1988robustness}, which generalizes normal-form perfect equilibrium by allowing trembles by a player’s opponents to be correlated.}
This differs from \citeapos{Myerson1986ACE} acceptable correlated equilibrium (ACE), which requires perturbations involving larger groups of players to be systematically less likely than those involving smaller groups. \citeapos{DM1996} perfect direct correlated equilibrium (PDCE) further requires these perturbations to be independent amongst players.

Second, CPE is invariant to the addition or removal of payoff-identical strategies. Because perturbations are placed directly on the distribution, adding or deleting duplicate strategies does not affect  CPE. This property is desirable because it ensures that a solution concept depends only on the game’s strategic structure, rather than on a particular representation \citep[cf.,][]{KM1986}. As we illustrate later in Section~\ref{sec:examples}, PDCE does not have this invariance property \citep[see also][]{CG2006}.

Third, in a CPE, all players derive their off-equilibrium beliefs from the same perturbed distribution over recommendation profiles. Consequently, conditional on the same event, it does not generate divergent posterior beliefs across players (see discussion in Example~\ref{eg:ours}). This distinguishes CPE from ACE, PDCE, as well as from correlated equilibrium with message-dependent trembles (CE-MDT) \citep*{luo2022revelation}, which explicitly allows players to hold subjective beliefs. 

The rest of the paper is organized as follows. We next discuss the related literature on refinements of correlated equilibrium. Section~\ref{sec:preliminary} introduces CPE and establishes its basic properties. Section~\ref{sec:results} provides a dual characterization of CPE and describes an iterative procedure for identifying the set of correlated perfect equilibria. 
Section~\ref{sec:examples} provides two examples showing that there is no set inclusion relationship between CPE and either ACE or PDCE. All proofs are provided in the Appendix.

\subsection*{Related Literature}
The refinement literature%
\footnote{There is another strand of literature that generalizes, rather than refines, correlated equilibrium, see, e.g., \cite{moulin1978strategically, grant2022delegation}.} 
on correlated equilibrium has primarily focused on perturbations of players' actions. The first such refinement is \cite{Myerson1986ACE}'s ACE, which allows players to make correlated mistakes under the restriction that mistakes involving larger groups of deviators are much less likely than those involving smaller groups. However, because players are not informed of their own trembles, in ACE, agents may form beliefs about others' mistakes that are inconsistent with the information available to them. This concern is raised by \citet{DM1996}. In response, they propose the notion of \emph{perfect correlated equilibrium} (PCE), defined as a perfect equilibrium of the extended communication game. In this framework, they impose independence across players’ mistakes, which in turn resolves the issue of inconsistent beliefs. Moreover, \citeauthor{DM1996} show that every PCE is also an ACE.

A drawback of PCE, however, is that it fails to satisfy the revelation principle: restricting PCEs to direct mechanisms, where messages coincide with players' strategies, comes with a loss of generality. A general characterization of PCEs therefore requires specifying an arbitrary message space. This complicates the use of PCE in applications. To address this challenge, \citet{luo2022revelation} introduce CE-MDT, which allows players to hold subjective, message-dependent beliefs about others’ mistakes. Unlike PCE, CE-MDT satisfies the revelation principle: every CE-MDT distribution can be implemented by the direct mechanism. Their framework is therefore more general in the message structures it allows; it nests PCE as a special case and coincides with a weak version of ACE (see their Proposition~3).

We view our approach as complementary to that of \citet{luo2022revelation}. Rather than restoring the revelation principle for general communication mechanisms, we take the direct mechanism as primitive and ask what perfection means when off-equilibrium beliefs arise from a common perturbation rather than player-specific trembles. In this sense, CPE captures a common-source perturbation model within the direct mechanism. Players do not need to hold heterogeneous, message-dependent subjective beliefs; rather, they derive their beliefs from the same perturbed distribution over recommendation profiles. Under the restriction that all players in CE-MDT share the \emph{same} subjective belief about correlated trembles over action recommendations, our notion coincides with CE-MDT.%
\footnote{Note that this comparison is within the direct mechanism, where messages coincide with action recommendations. \cite{luo2022revelation} made the same observation in their Section~5.3 for an earlier version of our paper. They also show in Section~5.1 that such a common-perturbation restriction may lead to a failure of the revelation principle.}

\section{Preliminaries} \label{sec:preliminary}
Consider a game $G=(N, (S_i)_{i\in N}, (u_i)_{i\in N})$, where $N$ is a finite set of players, $S_i$ is a finite set of pure strategies of player $i$, $S = \prod_{i \in N}S_i$ is the set of pure strategy profiles, and the function $u_i: S \to \mathbb{R}$ gives player $i$'s payoff at each strategy profile $s \in S$. 

A correlated strategy $\rho$ is a probability distribution over $S$ and we say that it is \emph{completely mixed} if it assigns positive probability to each element of $S$. Following \cite{Aumann1974, Aumann1987}, a correlated strategy $\rho$ is a \emph{correlated equilibrium} if for each player $i \in N$ and for every strategy $s_i\in S_i$, we have
\begin{equation} \label{eq:CE}
\sum_{s_{-i} \in S_{-i}} \rho(s_i, s_{-i}) \cdot u_i(s_i, s_{-i}) \geq \sum_{s_{-i} \in S_{-i}} \rho(s_i, s_{-i}) \cdot u_i(t_i, s_{-i}) \quad \text{for all~} t_i \in S_i.
\end{equation}
A correlated equilibrium $\rho$ is usually interpreted as follows: a mediator draws a strategy profile $s$ according to $\rho$ and privately recommends to each player $i$ her component $s_i$. Condition \eqref{eq:CE} requires that, given the distribution $\rho$ and assuming all opponents follow their recommendations, each player finds it optimal to follow her own recommendation.

It follows directly from \eqref{eq:CE} that the set of correlated equilibria is a compact, convex subset
of $\Delta S$ that contains all Nash equilibria. The existence of correlated equilibria follows from that of a Nash equilibrium \citep{nash1950non}.

\subsection*{Correlated Perfect Equilibrium}
We now introduce our notion of perfection for correlated equilibrium. In analogy to perfect equilibrium \citep{Selten1975}, where players may independently make arbitrarily small mistakes, we instead perturb the mediator's distribution over recommendations.

Formally, for a correlated strategy $\rho$, let $\rho_i(s_i)$ denote the unconditional probability that the mediator recommends $s_i$:
\begin{equation}\label{eq:rec_prob}
     \rho_i(s_i) := \sum_{s_{-i}\in S_{-i}} \rho(s_i, s_{-i}) \text{.}
\end{equation}
Thus, $\rho_i(s_i)$ is the ex-ante probability that player $i$ receives recommendation $s_i$. We write $S^\rho_i = \{s_i \in S_i: \rho_i(s_i)>0\}$ for the set of $i$'s strategies that lie in the support of $\rho$, and $S^\rho = \prod_{i \in N} S^\rho_i$ for the product of these sets across all players. Note that $S^\rho$ 
is the smallest product set of strategy profiles containing the support of $\rho$.

\begin{definition}[Correlated Perfect Equilibrium] \label{def:CPE}
   A correlated strategy $\rho \in \Delta S$ is a \emph{correlated perfect equilibrium} if there exists a sequence of completely mixed correlated strategies $(\rho^k)$ converging to $\rho$ such that, for each player $i \in N$ and every strategy $s_i \in S^\rho_i$, we have
\begin{equation} \label{eq:CPE}
 \sum_{s_{-i} \in S_{-i}} \rho^k(s_i, s_{-i}) \cdot u_i(s_i, s_{-i}) \geq \sum_{s_{-i} \in S_{-i}} \rho^k(s_i, s_{-i}) \cdot u_i(t_i, s_{-i}) \quad \text{for all~} t_i \in S_i, k \in \mathbb{N}.
\end{equation} 
\end{definition}

Clearly, every correlated perfect equilibrium is a correlated equilibrium, so our notion is a refinement. Intuitively, a correlated perfect equilibrium is a correlated equilibrium that is robust to arbitrarily small and possibly correlated noise near the equilibrium.

As with every equilibrium concept involving interactive decision making (e.g., Nash equilibrium), CPE has two layers of interpretation. The first describes players' actions, whereas the second reflects their beliefs about the behavior of others. In equilibrium (i.e., at the limit), these two layers coincide: players' beliefs correctly anticipate others' behavior. Under the first interpretation, condition \eqref{eq:CPE} imposes rationality only for strategies recommended with positive probability under $\rho$. Under the second interpretation, the converging sequence of completely mixed distributions $(\rho^k)$ pins down players' off-equilibrium beliefs about the behavior of others. The behavior attributed to others under these beliefs, however, may be suboptimal.\footnote{Indeed, after observing an event that occurs with zero probability in equilibrium, a player may entertain the possibility that her opponents are behaving suboptimally \citep[see][]{Reny1992}.} 

We remain agnostic about the source of the perturbations, requiring only that a single common sequence generate the off-equilibrium beliefs of all players. Depending on the application, one may interpret these perturbations as reflecting beliefs about some combination of mistakes by the mediator and players. 

We note two immediate implications of the definition. First, any correlated equilibrium that is completely mixed is also correlated perfect. Second, just as correlated equilibrium generalizes Nash equilibrium, our notion naturally extends \citet{Selten1975}’s perfect equilibrium, as we show below.  
\begin{lemma}
    \label{prop:existence}
Every perfect equilibrium is a correlated perfect equilibrium. 
\end{lemma} 
To see why this lemma holds, recall that a perfect equilibrium is a Nash equilibrium, and hence a correlated equilibrium. Its supporting sequence\textemdash which is a sequence of independently randomized, completely mixed strategy profiles \citep[see][]{Myerson1978Proper}\textemdash pins down the off-equilibrium beliefs and therefore also supports the equilibrium as a CPE. As the set of perfect equilibria is nonempty \citep{Selten1975}, an immediate consequence of Lemma~\ref{prop:existence} is the following.
\begin{corollary}
  There exists a correlated perfect equilibrium in any finite normal-form game.
\end{corollary}
Beyond existence, correlated perfection also inherits another desirable property of perfect equilibrium: it rules out weakly dominated strategies. 

\begin{lemma}\label{lem:weak}
In any correlated perfect equilibrium, no weakly dominated strategy is assigned positive probability.
\end{lemma}
Recall that, just as with Nash equilibrium, a correlated equilibrium may assign positive probability to weakly dominated strategies. Thus, Lemma~\ref{lem:weak} implies that the set of correlated perfect equilibria is contained within the set of correlated equilibria that assign zero probability to all weakly dominated strategies. 

\section{A Dual Representation of Correlated Perfect Equilibrium} \label{sec:results}
In this section, we provide a characterization of the set of correlated perfect equilibria. By definition, verifying that a correlated equilibrium is correlated perfect requires finding only a single supporting sequence that satisfies condition \eqref{eq:CPE}. However, ruling out correlated perfection is more demanding, as one must show that no such sequence exists. To address this challenge, we introduce a dual representation. This characterization provides not only an alternative interpretation of correlated perfection but also a systematic method for identifying the entire set of these equilibria.

To formalize this, we first introduce the notion of dual vectors \citep{NauMcCardle1990, Myerson1997Dual}. For each player $i$, consider a vector 
\[
\alpha_i = (\alpha_i\cBr{t_i}{s_i})_{s_i, t_i \in S_i} \in \R_+^{\left| S_i \times S_i \right|}
\] 
such that for each $s_i \in S_i$,
\[
\sum_{t_i \in S_i} \alpha_i\cBr{t_i}{s_i}=1\text{.}
\]
Following \cite{NauMcCardle1990}, the vector $\alpha_i$ can be interpreted as a randomized deviation plan for player $i$: if the mediator recommends strategy $s_i$, player $i$ deviates to $t_i$ with probability $\alpha_i\cBr{t_i}{s_i}$. For any strategy profile $s$, the expected gain to player $i$ from following deviation plan $\alpha_i$ is 
\[
D_i(s, \alpha_i) := \sum_{t_i \in S_i} \alpha_i\cBr{t_i}{s_i} \big( u_i(t_i, s_{-i}) - u_i(s_i, s_{-i})\big).
\]
\begin{definition}[Dual vector] \label{def:dual_vec}
    A profile of deviation plans $\alpha= (\alpha_i)_{i \in N}$ is a \emph{dual vector} if for all $s \in S$,  the sum of all players' expected gains from their respective deviations is nonnegative:
\begin{equation} \label{eq:dual}  
    \sum_{i \in N} D_i(s, \alpha_i) \geq 0.
\end{equation}
\end{definition}
In other words, a dual vector specifies unilateral deviation plans that induce a weak aggregate utility gain for any strategy profile $s$.

For a given correlated equilibrium $\rho$, recall that $S^\rho$ denotes the smallest product set containing the support of $\rho$. For any strategy $s_i$ of player~$i$, let $\delta_{s_i}$ denote the degenerate distribution that assigns probability one to $s_i$. 

\begin{definition}[Restricted dual vector]
A dual vector $\alpha = (\alpha_i)_{i \in N}$ is said to be \emph{$S^\rho$-restricted} if, for every player $i\in N$, 
\[
  \alpha_i(\cdot\mid s_i)=\delta_{s_i}\qquad \text{for all } s_i\notin S_i^\rho\text{.}
\]
\end{definition}
In words, for every player $i$, an $S^\rho$-restricted deviation plan $\alpha_i$ always prescribes that the agent follow the mediator's recommendation whenever the recommended strategy $s_i$ is \textit{not} in $S^\rho_i$. Thus, the deviation plans are restricted to $S^\rho$ in the sense that they do not specify deviations from strategies outside of $S_i^\rho$. It is worth emphasizing that, for each player $i$, a deviation plan $\alpha_i$ is fixed conditional on player $i$'s own recommended strategy $s_i$, not on the full strategy profile $s$. Thus, if two distinct full strategy profiles recommend the same $s_i$ to player $i$, the same deviation plan must be used in both cases. 

We are ready to state our main result. 

\begin{theorem} \label{thm:dual}
A correlated equilibrium $\rho$ is correlated perfect if and only if for every $S^\rho$-restricted dual vector $\alpha$ and every $s \in S$, \eqref{eq:dual} holds with equality.
\end{theorem}
Theorem \ref{thm:dual} shows that whether or not a correlated equilibrium $\rho$ is correlated perfect depends only on its minimal product support $S^\rho$. A direct implication is that if two correlated equilibria $\rho$ and $\rho'$ share the same product support ($S^\rho = S^{\rho'}$), then either both are correlated perfect or neither is. Moreover, since the set of correlated equilibria supported on any fixed product set of strategies forms a convex polyhedron, the set of correlated perfect equilibria can be expressed as a finite union of convex polyhedra. However, a union of convex sets is not necessarily convex: as we will see in the example at the end of this section, a mixture of two correlated perfect equilibria can fail to be correlated perfect.

Recall that a distribution over strategy profiles $\rho \in \Delta S$ is a correlated equilibrium if and only if, for any unrestricted dual vector, \eqref{eq:dual} holds with equality whenever $\rho(s)>0$
(see \citealp[Proposition~2]{NauMcCardle1990}; \citealp[Section~3]{Myerson1997Dual}). Intuitively, at every strategy profile in the support of a correlated equilibrium $\rho$, the aggregate utility gain from any collection of unilateral deviation plans is exactly zero. Theorem~\ref{thm:dual} strengthens this condition by requiring it to hold at every strategy profile in the \emph{smallest product support} of $\rho$. In other words, correlated perfection requires that the same aggregate utility gain condition hold whenever each player is recommended a strategy used with positive probability under $\rho$, even if that particular strategy profile is never realized. 

Finally, observe that an $S^\rho$-restricted dual vector $\alpha$ is also an $S'$-restricted dual vector whenever $S^\rho \subseteq S'$. As a consequence, if the equality in Theorem \ref{thm:dual} holds for $S'$, it also holds for every product subset of $S'$. Thus, if a correlated equilibrium $\rho$ is not correlated perfect, then by Theorem \ref{thm:dual} any correlated equilibrium $\rho'$ with $S^{\rho'} \supseteq S^\rho$ is also not correlated perfect. This observation suggests an iterative procedure for computing the set of correlated perfect equilibria: begin with the entire strategy space $S$ and check whether condition \eqref{eq:dual} holds with equality for all (unrestricted) dual vectors and all $s \in S$. If so, then every correlated equilibrium is correlated perfect. Otherwise, restrict attention to a smaller product set $S'$ and repeat the previous procedure. Once a correlated perfect equilibrium $\rho$ with $S^\rho = S'$ is identified, every correlated equilibrium whose product support is contained in $S'$ is also correlated perfect. The process terminates once every remaining product set is contained in one that is already associated with a correlated perfect equilibrium.

\subsection*{An Illustrative Example}

\begin{figure}[htbp]
\centering
\renewcommand{\arraystretch}{1.15}
\[
\begin{array}{c|c|c|c|}
\multicolumn{1}{c}{} 
& \multicolumn{1}{c}{x_2} 
& \multicolumn{1}{c}{y_2} 
& \multicolumn{1}{c}{z_2} \\ \cline{2-4}
x_1 & 0,0 & 0,0 & 0,0 \\ \cline{2-4}
y_1 & 0,0 & 1,-2 & -2,1 \\ \cline{2-4}
z_1 & 0,0 & -2,1 & 1,-2 \\ \cline{2-4}
\end{array}
\]
\caption{A two-player game from \cite{DM1996}.}
\label{fig:dm-game}
\end{figure}

We now illustrate the iterative process for finding the set of CPEs using a two-person game in Figure~\ref{fig:dm-game}, taken from \citet[p.~295]{DM1996}, where $S_i = \{x_i, y_i, z_i\}$ for $i =1, 2$. Let $\rho_{ab}$ denote the probability assigned to the profile $(a,b)$ with $a\in S_1$ and $b \in S_2$. First, observe that the set of correlated equilibria consists of probability distributions supported on $\{(x_1,x_2), (x_1,y_2),(x_1,z_2),(y_1,x_2),(z_1,x_2)\},$
satisfying $\frac12\rho_{xz}\leq \rho_{xy}\leq 2\rho_{xz}
$ and $\frac12\rho_{zx}\leq \rho_{yx}\leq 2\rho_{zx}.$

We begin with the full product set $S=S_1\times S_2$. Consider the unrestricted deviation plan under which, for every $i=1,2,$ and $s_i\in S_i$, $\bar\alpha_i(\cdot\mid s_i)=\delta_{x_i}$. That is, each player switches to $x_i$ after every recommendation. It is easy to verify that this is a dual vector; in particular, condition~\eqref{eq:dual} holds strictly for every $s\in\{y_1,z_1\}\times\{y_2,z_2\}$. Hence, according to the iterative process, we move to smaller product sets. Notice that the constraints from correlated equilibrium imply that \(\rho_{xy} >0\) iff  \(\rho_{xz}>0\), and \(\rho_{yx} >0\) iff \(\rho_{zx}>0\). Thus, after ruling out equilibria with \(S^\rho=S\), any remaining candidate must have either $\rho_{xy}=\rho_{xz}=0$ or $\rho_{yx}=\rho_{zx}=0$. 

Consider the product support $S'= \{x_1\}\times S_2$. Let $\alpha$ be a $S'$-restricted dual vector and write  
$a_y= \alpha_1(y_1\mid x_1)$ and $
a_z=\alpha_1(z_1\mid x_1)$. 
Since player 2's payoff is zero whenever player 1 plays $x_1$, condition~\eqref{eq:dual} at $(x_1,y_2)$ and $(x_1,z_2)$ implies $a_y-2a_z\geq 0$ and 
$-2a_y+a_z\geq 0$.
Hence $a_y=a_z=0$, so player 1 must follow the recommendation $x_1$. Applying condition~\eqref{eq:dual} to the remaining relevant profiles similarly implies that player 2 must follow each recommendation in $S_2$. Thus every $S'$-restricted dual vector prescribes no deviation from any recommendation, and condition~\eqref{eq:dual} holds with equality at every profile. The case for the product support $S^{''} = S_1\times\{x_2\}$ holds by symmetry. We therefore conclude that the set of CPEs is the union of two convex sets: the subset of correlated equilibria such that $\rho_{yx} = \rho_{zx} = 0$ and the subset of correlated equilibria such that $\rho_{xy} = \rho_{xz} = 0$. 

This example also shows that the set of CPEs need not be convex. For instance, the distribution assigning probability $1/2$ to each of $(x_1, y_2)$ and $(x_1, z_2)$ is a CPE, as is the distribution assigning probability $1/2$ to each of $(y_1, x_2)$ and $(z_1, x_2)$. However, their average assigns probability 1/4 to each of $(x_1, y_2)$, $(x_1, z_2)$, $(y_1, x_2)$, and $(z_1, x_2)$, and has full product support. Hence, by the argument above, it is not a CPE. 

\section{Examples} \label{sec:examples}
In this section, we present two additional examples that clarify the relationship between CPE and two refinements of correlated equilibrium: acceptable correlated equilibrium (ACE) \citep{Myerson1986ACE} and perfect direct correlated equilibrium (PDCE) \citep{DM1996}. 
Before turning to the examples, we recall the definitions of these two refinements. 

Formally, for any set of players $D \subseteq N$, let $S_D =  \prod_{i \in D} S_i$ with the convention that $S_{\emptyset}=\{\emptyset\}$. For any $s \in S$ and $s'_D \in S_D$, let $(s_{-D}, s'_D)$ denote the strategy profile in which player $i$ chooses $s'_i$ if $i \in D$ and $s_i$ otherwise. 

\begin{definition}[Acceptable Correlated Equilibrium; \cite{Myerson1986ACE}] \label{def:ACE}
For some $\eps \in(0,1)$, a distribution $\eta^{\eps} \in \Delta \bigl( S \times \left(\bigcup_{D \subseteq N} S_D \right) \bigr)$ is an \emph{$\eps$-correlated equilibrium} if it satisfies the following three conditions:
   \begin{compactenum}[(i)]
     \item  For each  $s \in S$, $i \in N$, $D \subseteq N \setminus \set{i}$, and each $s'_D \in S_D$,
     \[
     \eps \eta^{\eps} (s, s'_D) \geq (1-\eps) \sum_{s'_i \in S_i} \eta^{\eps}(s, s'_{D \cup \set{i}}) \text{;}
     \]
     \item For each $s \in S$, $i \in N$,  $D \subseteq N \setminus \set{i}$, and each $s'_{D \cup \set{i}} \in S_{D \cup \set{i}}$,
     \[
      \text{if}\quad \eta^{\eps}(s, s'_D) >0, \quad\text{then}\quad \eta^{\eps}(s, s'_{D \cup \set{i}}) >0  \text{;} 
     \]
     \item For each $i \in N$,  $s_i \in S_i$, 
     \[
       \sum_{s_{-i} \in S_{-i}} \sum_{D \subseteq N \setminus \set{i}} \sum_{s'_D \in S_D}
         \eta^{\eps}(s, s'_D) \bigl[ u_i(s_{-D}, s'_D) - u_i(s_{-(D \cup \set{i})}, s'_{D \cup \set{i}}) \bigr]
       \geq 0 \text{~for all~} s_i' \in S_i.
     \]
   \end{compactenum}  
A distribution $\rho$ over $S$ is an \emph{acceptable correlated equilibrium} if there exists a sequence of $\eps$-correlated equilibria $\eta^\eps$ such that $\lim_{\eps \to 0}\eta^\eps(s, \emptyset) = \rho(s)$ for every $s \in S$.
\end{definition}
In words, $\eta^{\eps}(s, s'_D)$ represents the probability that the mediator draws and recommends the profile $s$, the players in $D$ tremble to $s'_D$, and the profile $(s_{-D},s'_{D})$ is ultimately played. Hence, $\eta^{\eps}(s, \emptyset)$ is the probability that the mediator recommends $s$ and no player trembles. As with a CPE,  establishing that a distribution $\rho$ is an ACE requires finding a sequence $\eta^{\eps}$ that converges to $\rho$ and satisfies conditions (i)--(iii). Essentially, condition (i) requires that trembles involving larger subsets of players be less likely than those involving smaller subsets, whereas condition (ii) requires every possible tremble to have positive conditional probability. Condition (iii) requires that, conditional on being recommended $s_i$ and not trembling herself, player~$i$ find it optimal to follow $s_i$ against arbitrarily small trembles by the other players.

We next turn to PDCE. As discussed above, \citet{DM1996} define PCE using general communication devices, whereas CPE is formulated for the canonical direct mechanism. We therefore focus on PDCE, the PCE induced by a direct mechanism, to facilitate the comparison with CPE. 

For $k \in \mathbb{N}$, let $\sigma_i^k\cBr{s'_i}{s_i}$ denote the probability that player~$i$ plays $s'_i$ conditional on being recommended $s_i$, and let $\sigma^k = \bigl(\sigma_i^k\cBr{\cdot}{s_i}\bigr)_{s_i \in S_i,\,i \in N}$ denote a profile of conditional strategies. For a distribution $\rho \in \Delta S$ and $s \in S$, define
\[
 \rho^{\sigma^k}_{i}(s_i, s_{-i}) \;=\; \sum_{s'_{-i}\in S_{-i}} \rho(s_i, s'_{-i})
        \prod_{j \ne i} \sigma_j^k\cBr{s_j}{s'_j}.
\]
From player $i$’s perspective conditional on receiving recommendation $s_i$ and not trembling herself, this expression is proportional to the probability that the realized strategy profile is $s$ when the mediator draws recommendations according to $\rho$ and the other players independently tremble according to $\sigma^k$. Recall that $S^\rho_i$ is the set of player $i$'s strategies that lie in the support of $\rho$. 

\begin{definition}[Perfect Direct Correlated Equilibrium; \cite{DM1996}] \label{def:PDCE}
A distribution $\rho \in \Delta S$ is a \emph{perfect direct correlated equilibrium} (PDCE) if there exists a sequence of completely mixed conditional strategy profiles $(\sigma^k)$ such that:
\begin{compactenum}
    \item[(i)] For each $i \in N$ and $s_i \in S_i^{\rho}$, $\sigma_i^k\cBr{s_i}{s_i} \to 1$ as $k \to \infty$; and
    \item[(ii)] For every $i \in N$ and $s_i \in S_i^{\rho}$,
    \begin{equation*}
     \sum_{s_{-i}\in S_{-i}}  \rho^{\sigma^k}_{i}(s_i, s_{-i}) \cdot u_i(s_i, s_{-i}) \geq    \sum_{s_{-i}\in S_{-i}}  \rho_{i}^{\sigma^k}(s_i, s_{-i}) \cdot  u_i(s_i', s_{-i}),
        \quad \text{for all } s_i' \in S_i,~k \in \mathbb{N}.
    \end{equation*}
\end{compactenum}
\end{definition}
That is, in a PDCE, upon receiving recommendation $s_i$ and conditional on not trembling, each player $i$ finds it optimal to follow $s_i$ against small independent trembles by the opponents. 

\subsection*{Two Examples}
With these definitions in place, we begin by revisiting the two-person game in Figure \ref{fig:dm-game}. In this game, every correlated equilibrium is an ACE. This follows from \citet[Proposition 4]{DM1996}, which shows that in any two-person game, ACEs are those correlated equilibria that never assign positive probability to weakly dominated strategies. Meanwhile, the unique PDCE is $\rho_{x x} = 1$ \citep[see][Proposition 10]{DM1996}. Thus, in this particular two-person game, every PDCE is a CPE, and every CPE is an ACE.%
\footnote{Generally, every PDCE is an ACE \citep[see][Proposition~3]{DM1996}.}
Nonetheless, the (proper) set inclusions between CPE and the other refinements do not hold in general, as we will show below.

\begin{example}
\label{eg:Myerson}
\begin{figure}[htbp]
\centering
\begin{subfigure}{\textwidth}
  \centering
  \begin{tabularx}{0.75\textwidth}{ >{$}c<{$}| *{3}{ >{\centering\arraybackslash$}X<{$} |} c >{$}c<{$}| *{3}{>{\centering\arraybackslash$}X<{$} |} }
    \multicolumn{1}{c}{} & \multicolumn{1}{c}{$x_2$} & \multicolumn{1}{c}{$y_2$} & \multicolumn{1}{c}{$z_2$} &
    \qquad &
    \multicolumn{1}{c}{} & \multicolumn{1}{c}{$x_2$} & \multicolumn{1}{c}{$y_2$} & \multicolumn{1}{c}{$z_2$} 
    \\ \cline{2-4} \cline{7-9}
    x_1	& 1,1,2	& 2,0,0	& 2,0,0 	&&	x_1	& 3,3,1	& 3,3,1	& 3,3,1 
    \\ \cline{2-4} \cline{7-9}
    y_1	& 0,2,0	& 3,0,0	& 0,3,0 	&&	y_1	& 3,3,1	& 3,3,1	& 3,3,1 
    \\ \cline{2-4} \cline{7-9}
    z_1	& 0,2,0	& 0,3,0	& 3,0,0 	&&	z_1	& 3,3,1	& 3,3,1	& 3,3,1 
    \\ \cline{2-4} \cline{7-9}
    \multicolumn{9}{c}{} \\[-9pt]
    \multicolumn{1}{c}{} & \multicolumn{3}{c}{$x_3$} &
    &
    \multicolumn{1}{c}{} & \multicolumn{3}{c}{$y_3$}
  \end{tabularx}
  \caption{\citeapos{Myerson1986ACE} three-person game}
  \label{subfig:MyersonGame}  
\end{subfigure}

\vspace{4ex}

\begin{subfigure}{\textwidth}
  \centering
  \begin{tabularx}{0.75\textwidth}{ >{$}c<{$}| *{3}{ >{\centering\arraybackslash$}X<{$} |} c >{$}c<{$}| *{3}{>{\centering\arraybackslash$}X<{$} |} }
    \multicolumn{1}{c}{} & \multicolumn{1}{c}{$x_2$} & \multicolumn{1}{c}{$y_2$} & \multicolumn{1}{c}{$z_2$} &
    \qquad &
    \multicolumn{1}{c}{} & \multicolumn{1}{c}{$x_2$} & \multicolumn{1}{c}{$y_2$} & \multicolumn{1}{c}{$z_2$} 
    \\ \cline{2-4} \cline{7-9}
    x_1	& \eps	& \eps	& \eps 	&&	x_1	& \eps	& \eps	& \eps 
    \\ \cline{2-4} \cline{7-9}
    y_1	& \eps	& 3\eps	& \eps 	&&	y_1	& \eps	& 1		& \eps 
    \\ \cline{2-4} \cline{7-9}
    z_1	& \eps	& 7\eps	& \eps 	&&	z_1	& \eps	& \eps	& \eps 
    \\ \cline{2-4} \cline{7-9}
    \multicolumn{9}{c}{} \\[-9pt]
    \multicolumn{1}{c}{} & \multicolumn{3}{c}{$x_3$} &
    &
    \multicolumn{1}{c}{} & \multicolumn{3}{c}{$y_3$}
  \end{tabularx}
  \caption{A supporting sequence for $\delta_{(y_1, y_2, y_3)}$ to be correlated perfect.}
  \label{subfig:MyersonCPE}
\end{subfigure}
\caption{A CPE that is not an ACE.}
\label{fig:Myerson}
\end{figure} 
Consider a three-person game in Figure \ref{subfig:MyersonGame} where $S_1 = \{x_1, y_1, z_1\}$, $S_2 = \{x_2, y_2, z_2\}$ and $S_3 = \{x_3, y_3\}$. As shown by \citet[p.~143]{Myerson1986ACE}, 
the \emph{unique} acceptable correlated equilibrium in this game is the degenerate distribution that assigns probability one to strategy profile $(x_1,x_2,x_3)$. 

In contrast, we show that the degenerate distribution $\delta_{(y_1,y_2,y_3)}$, which assigns probability one to $(y_1,y_2,y_3)$, is a correlated perfect equilibrium using Definition \ref{def:CPE}. To this end, we construct a completely mixed sequence of correlated strategies that converges to $\delta_{(y_1, y_2, y_3)}$. This sequence is shown in Figure~\ref{subfig:MyersonCPE}.\footnote{For expositional simplicity, we omit the simple normalization needed to form a probability distribution.} For each $\eps>0$ in this sequence, it is straightforward to verify that each strategy recommended with positive probability in the limit is a best response for the corresponding player. For example, if player 1 is recommended $y_1$, her conditional expected payoffs from choosing $x_1, y_1$ and $z_1$ are $15\eps +3$, $15\eps +3$ and $9 \eps + 3$, respectively; thus following the recommendation $y_1$ is indeed optimal. An analogous calculation shows that $y_2$ is optimal for player $2$. For player $3$, $y_3$ yields a strictly higher expected payoff than $x_3$ and is therefore optimal. By definition, $\delta_{(y_1, y_2, y_3)}$ is a correlated perfect equilibrium. 
\end{example}

The distribution in Figure~\ref{subfig:MyersonCPE} supports $\delta_{(y_1,y_2,y_3)}$ as a CPE, but not as an ACE. This difference reflects the common-source interpretation of the perturbations discussed before. Although ACE permits correlated mistakes across players, it imposes a strict ordering on their probabilities (see condition (i) in Definition \ref{def:ACE}): deviations by larger groups of players must be infinitesimally less likely than deviations by smaller groups. The distribution in Figure~\ref{subfig:MyersonCPE} violates this requirement because it assigns the same order of probability to profiles with different numbers of deviators\textemdash for example, $(y_1,y_2,x_3)$ involves one deviator, player 3, whereas $(z_1,y_2,x_3)$ involves two deviators, players 1 and 3. 

By contrast, CPE imposes no such ordering: since it perturbs the distribution over recommendation profiles, a single ``noisy'' event at the source can result in correlated recommendation errors affecting multiple players. It is therefore possible for a multiple-player deviation to be more likely than a single-player deviation.

\begin{example}
\label{eg:ours}
\begin{figure}[htbp]
\centering
\begin{subfigure}{\textwidth}
  \centering
  \begin{tabularx}{\textwidth}{>{$}c<{$}| *{4}{>{\centering\arraybackslash$}X<{$}|} c >{$}c<{$}| *{4}{>{\centering\arraybackslash$}X<{$}|}}
    \multicolumn{1}{c}{} & \multicolumn{1}{c}{$w_2$} & \multicolumn{1}{c}{$x_2$} & \multicolumn{1}{c}{$y_2$} & \multicolumn{1}{c}{$z_2$} 
    &\qquad &
    \multicolumn{1}{c}{} & \multicolumn{1}{c}{$w_2$} & \multicolumn{1}{c}{$x_2$} & \multicolumn{1}{c}{$y_2$} & \multicolumn{1}{c}{$z_2$} \\ \cline{2-5} \cline{8-11}
    w_1	& 0,0,0	& 0,0,0	& 0,0,0	& 0,0,0	
    &&
    w_1	& 0,0,0	& 0,0,0	& 1,0,0	& 1,0,0
    \\ \cline{2-5} \cline{8-11}
    x_1	& 0,0,0	& 0,0,0	& 0,0,0	& 0,0,0	
    &&
    x_1	& 0,0,0	& 0,0,0	& 1,0,0	& 1,0,0
    \\ \cline{2-5} \cline{8-11}
    y_1	& 0,1,0	& 0,1,0	& 1,-2,0	& -2,1,0
    &&
    y_1	& 0,0,0	& 0,0,0	& 1,-2,0	& -2,1,0	
    \\ \cline{2-5} \cline{8-11}
    z_1	& 0,1,0	& 0,1,0	& -2,1,0	& 1,-2,0
    &&
    z_1	& 0,0,0	& 0,0,0	& -2,1,0	& 1,-2,0	
    \\ \cline{2-5} \cline{8-11}
    \multicolumn{11}{c}{} \\[-9pt]
    \multicolumn{1}{c}{} & \multicolumn{4}{c}{$x_3$} &
    &
    \multicolumn{1}{c}{} & \multicolumn{4}{c}{$y_3$}
  \end{tabularx}
  \caption{A three-person game}
  \label{subfig:ours}
\end{subfigure}

\vspace{4ex}

\begin{subfigure}{\textwidth}
  \centering
  \begin{tabularx}{\textwidth}{>{$}c<{$}| *{4}{>{\centering\arraybackslash$}X<{$}|} c >{$}c<{$}| *{4}{>{\centering\arraybackslash$}X<{$}|}}
    \multicolumn{1}{c}{} & \multicolumn{1}{c}{$w_2$} & \multicolumn{1}{c}{$x_2$} & \multicolumn{1}{c}{$y_2$} & \multicolumn{1}{c}{$z_2$} 
    &\qquad &
    \multicolumn{1}{c}{} & \multicolumn{1}{c}{$w_2$} & \multicolumn{1}{c}{$x_2$} & \multicolumn{1}{c}{$y_2$} & \multicolumn{1}{c}{$z_2$} \\ \cline{2-5} \cline{8-11}
    w_1	& 0		& 0		& 0	& 0	&& w_1	& 0	& 0	& 1/16	& 3/16
    \\ \cline{2-5} \cline{8-11}
    x_1	& 0		& 0		& 0	& 0	&& x_1	& 0	& 0	& 3/16	& 1/16
    \\ \cline{2-5} \cline{8-11}
    y_1	& 3/16	& 1/16	& 0	& 0	&& y_1	& 0	& 0	& 0		& 0	
    \\ \cline{2-5} \cline{8-11}
    z_1	& 1/16	& 3/16	& 0	& 0	&& z_1	& 0	& 0	& 0		& 0	
    \\ \cline{2-5} \cline{8-11}
    \multicolumn{11}{c}{} \\[-9pt]
    \multicolumn{1}{c}{} & \multicolumn{4}{c}{$x_3$} &
    &
    \multicolumn{1}{c}{} & \multicolumn{4}{c}{$y_3$}
  \end{tabularx}
  \caption{A perfect (direct) correlated equilibrium}
  \label{subfig:oursDM}
\end{subfigure}
\caption{A PDCE that is not a CPE.}
\label{fig:ours}
\end{figure}
Consider the three-person game in Figure~\ref{subfig:ours}, where $S_1=\{w_1, x_1, y_1, z_1\}$, $S_2=\{w_2, x_2, y_2, z_2\}$ and $S_3=\{x_3, y_3\}$. For simplicity, we set player~3's payoffs to zero. One can verify that the correlated strategy $\rho$ given in Figure~\ref{subfig:oursDM} is a PDCE (see \autoref{sec:omitted}).  
 
We now show that $\rho$ is not correlated perfect using Theorem \ref{thm:dual}. Since $S^\rho=S$, it suffices to exhibit an unrestricted dual vector $\alpha=(\alpha_1,\alpha_2,\alpha_3)$ that makes \eqref{eq:dual} strict on some profiles. Define
\begin{compactenum}
  \item[(i)] for $i=1,2$: $\alpha_i\cBr{\cdot}{w_i}=\alpha_i\cBr{\cdot}{y_i}=\alpha_i\cBr{\cdot}{z_i}=\delta_{w_i}$ and $\alpha_i\cBr{\cdot}{x_i}=\delta_{x_i}$;
  \item[(ii)] for $i=3$: $\alpha_3\!\cBr{\cdot}{x_3}=\delta_{x_3}$ and $\alpha_3\!\cBr{\cdot}{y_3}=\delta_{y_3}$.
\end{compactenum}
By construction, player~3 never deviates, so $D_3(s,\alpha_3)=0$ for all $s$. For players~1 and~2, the deviation plans send $y_i$ and $z_i$ to $w_i$ while leaving $w_i$ and $x_i$ unchanged. Hence, for any $s\in\{y_1,z_1\}\times\{y_2,z_2\}\times\{x_3,y_3\}$ we have
\[
\sum_{i=1}^3 D_i(s,\alpha_i)
=\big(u_1(w_1,s_{-1})-u_1(s)\big)+\big(u_2(w_2,s_{-2})-u_2(s)\big).
\]
For example, at $s=(y_1,y_2,x_3)$,
\[
D_1(s,\alpha_1)=0-1=-1,\qquad D_2(s,\alpha_2)=1-(-2)=+3,
\]
so $\sum_i D_i(s,\alpha_i)=2>0$. 

Similar calculations show that \eqref{eq:dual} holds with strict inequality for each $s\in\{y_1,z_1\}\times\{y_2,z_2\}\times S_3$. For $s\notin \{y_1,z_1\}\times\{y_2,z_2\}\times S_3$, either no deviation is prescribed or the payoff difference is zero, so \eqref{eq:dual} holds with equality. Thus $\alpha$ is an
unrestricted dual vector that yields a strict inequality on some $s \in S$, and by Theorem~\ref{thm:dual}, $\rho$ is not
correlated perfect. 
\end{example}

To further see why the PDCE in Figure \ref{subfig:oursDM} is not a CPE, recall that, in a PDCE, each player evaluates her recommendation conditional on \textit{not trembling} herself. 
When player 1 receives the recommendation  $y_1$ or $z_1$, he evaluates the optimality of his action conditional on player 2 playing either $y_2$ or $z_2$ since only those cases are payoff-relevant. If he is not trembling, the only recommendation profiles in the support of $\rho$ that are consistent with the recommendation he receives are $(s_1,w_2,x_3)$ and $(s_1,x_2,x_3)$ where $s_1 \in \{y_1, z_1\}$. Thus, player 1 believes that player~3 plays $x_3$. Likewise, when player 2 receives the recommendation  $y_2$ or $z_2$, she also evaluates the optimality of her action conditional on player 1 playing either $y_1$ or $z_1$. If she is not trembling, then the only recommendation profiles in the support of $\rho$ that are consistent with her recommendation lead her to believe that player~3 plays $y_3$. As a result, conditional on the same event that both players 1 and 2 play strategies in $\{y_i, z_i \}$, the two players hold different beliefs about player 3's action. This difference in beliefs is crucial for sustaining the PDCE, since each of them would have deviated to their $w$ or $x$ strategy had they held the other player's belief. 

In contrast, such player-specific rationalizations are incompatible with CPE: players' off-equilibrium beliefs are generated by a single common perturbing sequence rather than by player-specific trembles. Hence, players must hold the same belief about player 3's action, making it impossible for both players 1 and 2 to play, say, their $y$ strategies in the same CPE.

\subsection*{Duplication Invariance}
Notice that the first two strategies of players 1 and 2 in Example~\ref{eg:ours} are payoff-equivalent duplicates. This simplification is purely expositional: one could, for instance, slightly perturb player 3’s payoffs at $(w_1, w_2, x_3)$ and $(w_1,w_2,y_3)$ to make these strategies strategically distinct without changing the argument.

Indeed, CPE is duplication-invariant: adding or removing duplicate strategies does not affect whether the corresponding correlated equilibrium is correlated perfect. This follows directly from the definition of CPE, since duplicate strategies generate the same optimality constraints. As a result, such duplications have no effect on the existence of a supporting sequence. In Example~\ref{eg:ours}, if $w_i$ and $x_i$ are collapsed into a single strategy $a_i$ for each $i=1,2$, one could apply the same dual-vector argument by replacing $w_i$ and $x_i$ by $a_i$, and the projected correlated equilibrium still fails to be correlated perfect.

On the other hand, the PDCE in Example~\ref{eg:ours} is sensitive to the removal of duplicated strategies. If $w_i$ and $x_i$ are collapsed into a single strategy for each $i$, the projection of the distribution in Figure~\ref{subfig:oursDM} onto the resulting reduced game is no longer a PDCE.%
\footnote{See Appendix~\ref{sec:omitted} for details.} 
Recall that under a PDCE, mistakes are assumed to be independent for each player, conditional on the message she receives. By duplicating strategies, however, the mediator gains the ability to correlate the messages sent to players, thereby inducing correlated mistakes in realized actions and effectively circumventing the independence assumption. In fact, as shown in  \citet{CG2006}, given any finite normal-form game, every PCE---without restricting attention to the direct mechanism---can be implemented as a PDCE of a game obtained by duplicating players' pure strategies.

\section{Conclusion}
We propose a refinement of correlated equilibrium, which we call \emph{correlated perfect equilibrium} (CPE). In contrast to earlier refinements of correlated equilibrium that place perturbations on players' strategies, CPE places them on a single common object: the mediator's distribution over recommendation profiles.

This common-source formulation has several advantages. It permits correlated perturbations without imposing additional assumptions on their relative likelihoods, generates off-equilibrium beliefs from a common distribution without introducing additional player-specific private information, and is invariant to the duplication of payoff-identical pure strategies. As we illustrate through two examples, CPE differs from existing player-tremble refinements, such as ACE and PDCE, and there is no set-inclusion relationship between CPE and these refinements. Moreover, CPE never assigns positive probability to weakly dominated strategies and admits a dual characterization. This characterization also reveals its geometric structure---as a finite union of convex polyhedra---and provides an iterative procedure for identifying these equilibria.

As our objective is to analyze behavior when off-equilibrium beliefs are derived from a common source, we remain agnostic about the origin of the perturbations. They may reflect some combination of errors by the mediator and the players. For instance, the mediator may be authorized to select strategies on the players' behalf, subject to mistakes that occur with vanishing probability. Taking a stand on the source of the perturbations would require specifying the message space and players' behavior following off-equilibrium messages.  We leave this as a potentially fruitful direction for future work. 

Our agnostic approach to the origin of perturbations does not impose further likelihood restrictions among suboptimal strategies. In particular, a dominated strategy is not necessarily infinitely less likely than an undominated strategy that remains suboptimal at an equilibrium. Nonetheless, by providing a simple common source for off-equilibrium beliefs, CPE offers a foundation for further refinements of correlated equilibrium. For instance, in the spirit of strong belief in rationality \citep{battigalli2002strong}, one could require dominated strategies to be infinitely less likely than undominated ones along the perturbing sequence.

One may wonder whether there is a connection between CPE and dual reduction, as well as elementary games obtained through iterative dual reduction \citep{Myerson1997Dual, viossat2010properties}. This connection is limited, since there exist CPEs that are not preserved under iterative dual reduction.%
\footnote{To see this, recall that perfect equilibria are CPEs, but they do not necessarily survive iterative elimination of dominated strategies, which is a special form of iterative dual reduction. See, for example, the Outside Option Game \citep[Fig.~1]{vanDamme1989Stable}.} 
Thus, elementary games do not provide a full characterization of CPE. Nonetheless, every correlated equilibrium of an elementary game is a CPE of that elementary game because all dual vectors in an elementary game specify that every player follows the mediator’s recommendation with probability one. A natural question is whether, and under what conditions, correlated perfection is preserved when moving between a game and its dual reductions. Relatedly, it would be interesting to know whether CPE can be viewed as a perfect equilibrium of an auxiliary zero-sum game analogous to the one used by \citet{hart1989existence} to prove the existence of correlated equilibria. We leave these questions for future research.

\bibliographystyle{ecta}
\bibliography{reference}

\newpage
\appendix
\section{Omitted Proofs}
\begin{proof}[Proof of Lemma \ref{prop:existence}]
Let $\sigma \in \prod_{i \in N} \Delta S_i$ be a perfect equilibrium of the game $G$. 
By definition \citep{Selten1975, Myerson1978Proper}, there exists a sequence of completely (independently) mixed strategy profiles 
$(\sigma^k)$ converging to $\sigma$ such that, for each $k$, each player $i \in N$, and each 
$s_i \in S_i$ with $\sigma_i(s_i) > 0$, 
\[
\sum_{s_{-i} \in S_{-i}} \sigma_{-i}^k(s_{-i}) \, u_i(s_i, s_{-i})
   \;\;\geq\;\;
\sum_{s_{-i} \in S_{-i}} \sigma_{-i}^k(s_{-i}) \, u_i(t_i, s_{-i})
   \quad \text{for all } t_i \in S_i.
\]

Now construct a sequence of completely mixed correlated strategies $(\rho^k)$ by setting, 
for each $s \in S$, 
\[
\rho^k(s) := \prod_{i \in N} \sigma_i^k(s_i).
\]
Clearly, $\rho^k$ converges to $\sigma$. Moreover, multiplying both sides of the inequality above by 
$\sigma_i^k(s_i)$ yields condition~\eqref{eq:CPE}. Hence every perfect equilibrium is also 
a correlated perfect equilibrium.
\end{proof}

Recall that a pure strategy $s_i$ of player $i$ is said to be \emph{weakly dominated} if there exists a mixed strategy $\sigma_i \in \Delta S_i$ such that 
\[
u_i(s_i, s_{-i}) \leq \sum_{t_i \in S_i} \sigma_i(t_i) \cdot u_i(t_i, s_{-i}), \text{~for all~} s_{-i} \in S_{-i}, 
\]
and the inequality is strict for at least one $s_{-i} \in S_{-i}$. The following claim will be useful in proving Lemma~\ref{lem:weak} \citep[see also][Proposition~2]{weinstein2020best}. 

\begin{claim}\label{claim:dom}\citep[Theorem~1.7, pp.~30--31]{MyersonText}. A pure strategy $s_i \in S_i$ of player $i$ is weakly dominated if and only if, for each completely mixed correlated opponent strategy profile $\rho_{-i} \in \Delta S_{-i}$, there exists $t_i \in S_i$ such that
\[
 \sum_{s_{-i} \in S_{-i}} \rho_{-i}(s_{-i}) u_i(s_i, s_{-i}) <   \sum_{s_{-i} \in S_{-i}} \rho_{-i}(s_{-i}) u_i(t_i, s_{-i})\text{.}
\]
\end{claim}

\begin{proof}[Proof of Lemma \ref{lem:weak}]
Suppose toward a contradiction that $\rho$ is a correlated perfect equilibrium and that a weakly dominated strategy $s_i$ is in its support, i.e., $\rho_i(s_i)>0$. Since $\rho_i(s_i) >0$,  condition \eqref{eq:CPE} must hold for $s_i$ along some completely mixed sequence converging to $\rho$. But
by Claim \ref{claim:dom}, for every completely mixed correlated strategy $\hat \rho$, we have 
  $$  \sum_{s_{-i} \in S_{-i}} \hat\rho(s_i, s_{-i}) \cdot u_i(s_i, s_{-i}) < \sum_{s_{-i} \in S_{-i}}\hat \rho(s_i, s_{-i}) \cdot u_i(t_i, s_{-i}) \quad \text{for some~} t_i \in S_i. $$
That is, no sequence of completely mixed correlated strategies $\rho^k$ converging to $\rho$ can satisfy \eqref{eq:CPE} for $s_i$, a contradiction.  
\end{proof}

To prove Theorem \ref{thm:dual}, we will use the following version of Farkas' Lemma \citep[see, e.g.,][p.~89]{Schrijver1986}. 
\begin{lemma}[Farkas' Lemma]\label{lem:farkas}
 Let $A$ be an $m \times n$ matrix and $b$ be an $m$-length vector. Then exactly one of the following holds:
  \begin{compactenum}
    \item[(i)] There exists an $x \in \R^n$ such that $Ax \geq b$.
    \item[(ii)] There exists a $y \in \R_+^m$ such that $y^{\top}A =0$ and $y \cdot b > 0$.
 \end{compactenum}
\end{lemma}

\begin{proof}[Proof of Theorem \ref{thm:dual}]
We first show the only-if direction by contraposition. Let $\rho$ be a correlated equilibrium. Suppose there exists an $S^\rho$-restricted dual vector $\alpha$ and a profile $\bar s \in S$ such that 
\[
    \sum_{i\in N}\sum_{t_i\in S_i}\alpha_i\cBr{t_i}{\bar s_i}\bigl[u_i(t_i,\bar s_{-i})-u_i(\bar s_i,\bar s_{-i})\bigr] >0.
\]
Notice that some player~$i$ must be playing a strategy from $S^\rho_i$ at $\bar s$. Otherwise, since $\alpha$ is $S^\rho$-restricted, $\alpha_i\cBr{t_i}{\bar s_i}$ is nonzero only at $t_i=\bar s_i$ for every $i$, and every utility difference is zero.

For each $s \in S$, define
\[
G_\alpha(s) = \sum_{i \in N} \sum_{t_i \in S_i} \alpha_i(t_i \mid s_i)\bigl[ u_i(t_i, s_{-i})-u_i(s_i, s_{-i}) \bigr].
\]
Since $\alpha$ is a dual vector, $G_\alpha(s)  \ge 0$ for every $s \in S$. Moreover, by the strict inequality above, $G_\alpha(\bar s)>0$. Let $(\rho^k)$ be an arbitrary sequence of completely mixed correlated strategies converging to $\rho$. Since $\rho^k(\bar s) >0$, we have 
\[
\sum_{s \in S} \rho^k(s) G_\alpha(s)= \sum_{s \in S} \rho^k(s)   \sum_{i \in N} \sum_{t_i \in S_i} \alpha_i\cBr{t_i}{s_i} \bigl[ u_i(t_i, s_{-i})-u_i(s_i, s_{-i}) \bigr]
  > 0.
\]
Rearranging the sums:
\[
  \sum_{i \in N} 
  \sum_{s_i \in S_i} 
  \sum_{s_{-i} \in S_{-i}}
  \sum_{t_i \in S_i}
    \alpha_i\cBr{t_i}{s_i} \rho^k(s_i, s_{-i}) \bigl[ u_i(t_i, s_{-i})-u_i(s_i, s_{-i}) \bigr]
  > 0\text{.} 
\]
Since $\alpha$ is $S^\rho$-restricted, the terms for $s_i \notin S^\rho_i$ drop from the second sum. Hence we can write
\[
  \sum_{i \in N}
  \sum_{s_i \in S^\rho_i}
  \sum_{t_i \in S_i}
    \alpha_i\cBr{t_i}{s_i}  
    \sum_{s_{-i} \in S_{-i}} \rho^k(s_i, s_{-i}) \bigl[ u_i(t_i, s_{-i})-u_i(s_i, s_{-i}) \bigr]
  > 0\text{.}
\]
But this implies that there exist some $i$, $s_i \in S^\rho_i$ and $t_i \in S_i$ such that
\[
  \sum_{s_{-i} \in S_{-i}}\rho^k(s_i,s_{-i}) \bigl[u_i(t_i,s_{-i})-u_i(s_i,s_{-i})\bigr] > 0,
\]
which contradicts \eqref{eq:CPE}. As the choice of $\rho^k$ was arbitrary, this shows that every completely mixed sequence that converges to $\rho$ violates \eqref{eq:CPE}. Thus, no supporting sequence exists and we conclude that $\rho$ is not correlated perfect.

We now show the if-direction. Let $\rho$ be a correlated equilibrium. For easier exposition, adopt the shorthands $m':= \sum_{i \in N} \left(\left|S^\rho_i\right| \cdot \left|S_i\right| \right)$, $n=\left|S\right|$ and $m:= m' + n$. To apply Lemma~\ref{lem:farkas}, we will construct an $m \times n$ matrix $A$ as follows:
\begin{compactenum}
  \item[(a)] Each column corresponds to a strategy profile $s \in S$.
  \item[(b)] The first $m'$ rows are indexed by triples $(i,\tilde{s}_i, t_i)$ where $\tilde{s}_i \in S^\rho_i$ and $t_i \in S_i$. The entry at row $(i,\tilde{s}_i, t_i)$ and column $s=(s_i, s_{-i})$ is
  \[
    A\big[(i,\tilde s_i,t_i),\,s\big] \;=\;
    \begin{cases}
      u_i(\tilde{s}_i, s_{-i}) - u_i(t_i, s_{-i}), & \text{if } \tilde{s}_i = s_i\text{,}\\[4pt]
      0, & \text{otherwise.}
    \end{cases}
  \]
  \item[(c)] The last $n$ rows form an $n \times n$ identity matrix.
\end{compactenum}
Construct also a vector $b \in \R^m_+$ where the first $m'$ entries are $0$ and the last $n$ entries are $1$.

Now suppose for every $S^\rho$-restricted dual vector $\alpha=(\alpha_i)_{i\in N}$ and every $s\in S$, \eqref{eq:dual} holds with equality: 
\begin{equation} \label{eq:dualproof}
\sum_{i\in N} D_i(s, \alpha_i) = \sum_{i\in N}\sum_{t_i\in S_i}\alpha_i\cBr{t_i}{s_i}\bigl[u_i(t_i,s_{-i})-u_i(s_i,s_{-i})\bigr]=0.
\end{equation}
With the matrix $A$ and vector $b$ constructed, we will first argue that condition (ii) in Lemma~\ref{lem:farkas} is impossible and thus condition (i) must hold. We then use condition (i) to show that $\rho$ is a CPE. 

To this end, assume towards a contradiction that there exists $\hat{y} \in \R^m_+$ such that $\hat{y}^\top A =0$ and $\hat{y}\cdot b>0$. Denote a generic element in the first $m'$ entries of $\hat{y}$ by $\hat{\alpha}_i\cBr{t_i}{\tilde{s}_i}$, where $(i, \tilde{s}_i, t_i)$ is associated with the row of  matrix $A$. Denote a generic element in the last $n$ entries of $\hat{y}$ by $\hat{\beta}(s)$, where $s$ is associated with the identity row for column $s$ in matrix $A$. Given this notation, the $s$-component of $\hat{y}^\top A$ is
\begin{align}
  \left(\hat{y}^\top A\right)_s
  &= \sum_{i \in N} \sum_{\tilde{s}_i \in S^\rho_i} \sum_{t_i \in S_i}
      \hat{\alpha}_i\cBr{t_i}{\tilde{s}_i} A\big[(i,\tilde s_i,t_i),\,s\big] 
  + \hat{\beta}(s) \notag \\
  &= \sum_{i \in N} \sum_{t_i \in S_i} 
      \hat{\alpha}_i\cBr{t_i}{s_i} \bigl[u_i(s_i,s_{-i})-u_i(t_i,s_{-i})\bigr] 
  + \hat{\beta}(s) = 0 \text{,} \label{eq:dualsystem}
\end{align}
where the second equality follows because $A\big[(i,\tilde s_i,t_i),\,s\big]=0$ when $\tilde{s}_i \ne s_i$. We set $\hat \alpha_i(t_i \mid s_i) = 0$ whenever $s_i \not\in S^\rho_i$. Now, let $c$ be a constant such that
\[
  c= \left(1+\max_{i, \tilde{s}_i \in S^\rho_i}  \sum_{t_i \in S_i} \hat{\alpha}_i \cBr{t_i}{\tilde{s}_i} \right)^{-1} \text{.}
\]
Construct for each player~$i$, each $s_i \in S_i$,
\[
  \alpha_i \cBr{\cdot}{s_i} =
  \begin{cases}
    \big[1 - c \sum_{t_i \in S_i} \hat{\alpha}_i\cBr{t_i}{s_i}\big] \delta_{s_i} + c\, \hat{\alpha}_i \cBr{\cdot}{s_i}	&\text{if }s_i \in S_i^\rho\text{,} \\
    \delta_{s_i}	&\text{otherwise.}
  \end{cases}
\]
We claim that $\alpha = (\alpha_i)_{i \in N}$ is a $S^\rho$-restricted dual vector. Restriction to $S^\rho$ is obvious. Non-negativity of $\alpha$ is guaranteed by $\hat{y} \geq 0$ and the choice of $c$. It is also straightforward to verify that each $\alpha_i\cBr{\cdot}{s_i}$ is a probability distribution. Meanwhile, for each $s \in S$,
\begin{alignat*}{2}
  \sum_{i\in N} D_i(s, \alpha_i) &= \sum_{i\in N}\sum_{t_i\in S_i}\alpha_i\cBr{t_i}{s_i}\bigl[u_i(t_i,s_{-i})-u_i(s_i,s_{-i})\bigr] && \\
  &= \sum_{i \text{ s.t.\ }s_i \in S^\rho_i} \sum_{t_i \ne s_i} 
  c\, \hat{\alpha}_i \cBr{t_i}{s_i} \bigl[u_i(t_i,s_{-i})-u_i(s_i,s_{-i})\bigr] && \\
  &= c\, \hat{\beta}(s) \\
  & \geq 0\text{,} 
\end{alignat*}
where the third equality follows from \eqref{eq:dualsystem}. Hence $\alpha$ is an $S^\rho$-restricted dual vector. By \eqref{eq:dualproof}, the last weak inequality holds as equality. But this means $\hat{\beta}(s)=0$ for all $s \in S$. This contradicts $\hat{y} \cdot b >0$.

Thus, (ii) in Lemma~\ref{lem:farkas} cannot hold, so the primal system $Ax\ge b$ must be feasible. That is, there exists $\mu\in \R^{|S|}$
such that
\begin{align}
\sum_{s_{-i}\in S_{-i}}\mu(s_i,s_{-i})\bigl[u_i(s_i,s_{-i})-u_i(t_i,s_{-i})\bigr]&\ge 0
&&\forall\, i\in N,\ s_i\in S_i^\rho,\ t_i\in S_i,\label{eq:mu-br}\\
\mu(s)&\ge 1 &&\forall\, s\in S\text{.}\notag
\end{align}
We will use this $\mu$ to construct a supporting sequence for $\rho$ to be a CPE.

For each $k \in \mathbb{N}$, define
\[
\widehat\rho^k(s):=\rho(s)+\frac{1}{k}\cdot \mu(s) 
\]
and normalize to obtain
\[
\rho^k(s):=\frac{\widehat\rho^k(s)}{\sum_{\tau\in S}\widehat\rho^k(\tau)}\text{.}
\]
Then $\rho^k$ is completely mixed since $\mu(s)\ge 1$. Clearly, $\rho^k\to\rho$ as $k \to \infty$. It remains to show that \eqref{eq:CPE} holds for $\rho^{k}$. Fix any $i$, $ s_i \in S_i^\rho$ and $t_i \in S_i$. We can write 
\begin{align*}
  & \sum_{s_{-i} \in S_{-i}} \rho^{k}( s_i, s_{-i}) \big(u_i( s_i, s_{-i})-u_i(t_i, s_{-i})\big) \\
     = & \frac{1}{\sum_{s \in S} \hat{\rho}^k(s)}
  \Bigg(
    \sum_{s_{-i}\in S_{-i}} \rho(s) \big(u_i(s_i, s_{-i})-u_i(t_i, s_{-i})\big)
   + \frac{1}{k} \sum_{s_{-i}\in S_{-i}} \mu(s) \big(u_i(s_i, s_{-i})-u_i(t_i, s_{-i})\big)
  \Bigg) \text{.}
\end{align*} 
The first term is nonnegative since $\rho$ is a correlated equilibrium and the second term is nonnegative by \eqref{eq:mu-br}. Hence, Condition~\eqref{eq:CPE} holds along the sequence $\rho^k$. Therefore $\rho$ is a correlated perfect equilibrium.
\end{proof}

\section{Omitted Details in Example~\ref{eg:ours}} \label{sec:omitted}
To see that the correlated equilibrium $\rho$ in Figure~\ref{subfig:oursDM} is a PDCE, consider the following sequence of independent trembles. Throughout the rest of this section, we use $\eps>0$ as a shorthand for $\eps = 1/k$ where $k \in \mathbb{N}$ indexes the supporting sequence introduced in Definition~\ref{def:PDCE}. For each sufficiently small $\eps>0$, define
\begin{compactenum}
   \item[(i)] For $i= 1, 2$, 
\[
\sigma_i^\eps(s'_i\mid s_i)=
\begin{cases}
1-2\eps-\eps^2, & \text{if } s'_i=s_i\in\{w_i,x_i\},\\
1-3\eps,               & \text{if } s'_i=s_i\in\{y_i,z_i\},\\
\eps^2,                & \text{if } (s'_i,s_i)\in\{(z_i,w_i),(y_i,x_i)\},\\
\eps,                  & \text{otherwise.}
\end{cases}
\]
 \item[(ii)] For $i=3$, 
\[
\sigma_3^\eps(s_3'\mid s_3)=
\begin{cases}
1-\eps, & s_3'=s_3,\\
\eps,   & s_3'\neq s_3.
\end{cases}
\]
\end{compactenum}
Condition~(i) for PDCE is clearly satisfied. For Condition~(ii), consider player 1 and suppose the recommendation is $w_1$. Under $\rho$, the only opponent recommendations with positive probability are $(y_2,y_3)$ and $(z_2,y_3)$. Hence, for any opponent profile $(s_2,s_3)$,
\begin{align*}
\rho^1_{\sigma^\eps}(w_1, s_2, s_3) = \frac{1}{16} \cdot \sigma^\eps_2\cBr{s_2}{y_2} \cdot \sigma^\eps_3\cBr{s_3}{y_3} + \frac{3}{16}\cdot \sigma^\eps_2\cBr{s_2}{z_2} \cdot \sigma^\eps_3\cBr{s_3}{y_3}.
\end{align*}
From player 1's perspective, the game in Figure~\ref{subfig:ours} would be played according to Figure~\ref{fig:app_PDCE}.
 \begin{figure}[h]
\centering
  \begin{tabularx}{\textwidth}{>{$}c<{$}| *{4}{>{\centering\arraybackslash$}X<{$}|} c >{$}c<{$}| *{4}{>{\centering\arraybackslash$}X<{$}|}}
    \multicolumn{1}{c}{} & \multicolumn{1}{c}{$w_2$} & \multicolumn{1}{c}{$x_2$} & \multicolumn{1}{c}{$y_2$} & \multicolumn{1}{c}{$z_2$} 
    &\qquad &
    \multicolumn{1}{c}{} & \multicolumn{1}{c}{$w_2$} & \multicolumn{1}{c}{$x_2$} & \multicolumn{1}{c}{$y_2$} & \multicolumn{1}{c}{$z_2$} \\ \cline{2-5} \cline{8-11}
    w_1	& 	\frac{\eps^2}{4}	& \frac{\eps^2}{4}	& \frac{\eps}{16}	& \frac{3\eps}{16} - \frac{\eps^2}{2}	&& w_1	& \frac{\eps-\eps^2}{4} & \frac{\eps-\eps^2}{4}
     & \frac{1-\eps}{16}& \frac{3-11\eps+8\eps^2}{16}
    \\ \cline{2-5} \cline{8-11}
    x_1	& 0		& 0		& 0	& 0	&& x_1	& 0	& 0	& 0	& 0
    \\ \cline{2-5} \cline{8-11}
    y_1	& 0	& 0	& 0	& 0	&& y_1	& 0	& 0	& 0		& 0	
    \\ \cline{2-5} \cline{8-11}
    z_1	& 0	& 0	& 0	& 0	&& z_1	& 0	& 0	& 0		& 0	
    \\ \cline{2-5} \cline{8-11}
    \multicolumn{11}{c}{} \\[-11pt]
    \multicolumn{1}{c}{} & \multicolumn{4}{c}{$x_3$} &
    &
    \multicolumn{1}{c}{} & \multicolumn{4}{c}{$y_3$}
  \end{tabularx}
  \caption{Player 1's perspective of the game in Figure~\ref{subfig:ours}.}
\label{fig:app_PDCE}
 \end{figure}
The expected gain from deviating to $y_1$ is 
$$ \rho^{1}_{\sigma^\eps}(w_1,y_2,x_3)
-2\rho^{1}_{\sigma^\eps}(w_1,z_2,x_3) -3\rho^{1}_{\sigma^\eps}(w_1,z_2,y_3)= - \frac
{9}{16} + \frac{7}{4} \eps - \frac{1}{2}\eps^2.$$ 
 Likewise, the expected gain from deviating to $z_1$ is 
 \[
-2\rho^{1}_{\sigma^\eps}(w_1,y_2,x_3)
+\rho^{1}_{\sigma^\eps}(w_1,z_2,x_3) -3\rho^{1}_{\sigma^\eps}(w_1,y_2,y_3) = -\frac
{3}{16} + \frac{1}{4}\eps - \frac{1}{2}\eps^2.
 \]
 Both expressions are negative for sufficiently small $\eps>0$. 
 The expected gain from deviating to $x_1$ is zero since $w_1$ and $x_1$ are payoff-equivalent for player 1. Hence, no deviation is profitable for player 1 when recommended $w_1$. By analogous arguments, one can verify that player 1 has no incentive to deviate when recommended $x_1$, $y_1$, or $z_1$. By symmetry, the same arguments also hold for player 2. Finally, since player 3’s payoffs are constant across all outcomes, any recommendation is trivially optimal. Therefore, these trembles satisfy condition (ii) in the definition of PDCE, and we conclude that the distribution shown in Figure~\ref{subfig:oursDM} is a perfect direct correlated equilibrium.

\begin{figure}[htbp]
\centering
\begin{subfigure}{\textwidth}
  \centering
  \begin{tabularx}{0.75\textwidth}{>{$}c<{$}| *{3}{>{\centering\arraybackslash$}X<{$}|} c >{$}c<{$}| *{3}{>{\centering\arraybackslash$}X<{$}|}}
    \multicolumn{1}{c}{} & \multicolumn{1}{c}{$a_2$} & \multicolumn{1}{c}{$y_2$} & \multicolumn{1}{c}{$z_2$} 
    &\qquad &
    \multicolumn{1}{c}{} & \multicolumn{1}{c}{$a_2$} & \multicolumn{1}{c}{$y_2$} & \multicolumn{1}{c}{$z_2$} 
    \\ \cline{2-4} \cline{7-9}
    a_1 & 0,0,0 & 0,0,0 & 0,0,0
    &&
    a_1 & 0,0,0 & 1,0,0 & 1,0,0
    \\ \cline{2-4} \cline{7-9}
    y_1 & 0,1,0 & 1,-2,0 & -2,1,0
    &&
    y_1 & 0,0,0 & 1,-2,0 & -2,1,0
    \\ \cline{2-4} \cline{7-9}
    z_1 & 0,1,0 & -2,1,0 & 1,-2,0
    &&
    z_1 & 0,0,0 & -2,1,0 & 1,-2,0
    \\ \cline{2-4} \cline{7-9}
    \multicolumn{9}{c}{} \\[-9pt]
    \multicolumn{1}{c}{} & \multicolumn{3}{c}{$x_3$} &
    &
    \multicolumn{1}{c}{} & \multicolumn{3}{c}{$y_3$}
  \end{tabularx}
  \caption{The reduced game obtained by identifying $w_i$ and $x_i$ as $a_i$ for each $i=1,2$.}
  \label{subfig:ours-reduced}
\end{subfigure}

\vspace{4ex}

\begin{subfigure}{\textwidth}
  \centering
  \begin{tabularx}{0.75\textwidth}{>{$}c<{$}| *{3}{>{\centering\arraybackslash$}X<{$}|} c >{$}c<{$}| *{3}{>{\centering\arraybackslash$}X<{$}|}}
    \multicolumn{1}{c}{} & \multicolumn{1}{c}{$a_2$} & \multicolumn{1}{c}{$y_2$} & \multicolumn{1}{c}{$z_2$} 
    &\qquad &
    \multicolumn{1}{c}{} & \multicolumn{1}{c}{$a_2$} & \multicolumn{1}{c}{$y_2$} & \multicolumn{1}{c}{$z_2$} 
    \\ \cline{2-4} \cline{7-9}
    a_1 & 0 & 0 & 0
    &&
    a_1 & 0 & 1/4 & 1/4
    \\ \cline{2-4} \cline{7-9}
    y_1 & 1/4 & 0 & 0
    &&
    y_1 & 0 & 0 & 0
    \\ \cline{2-4} \cline{7-9}
    z_1 & 1/4 & 0 & 0
    &&
    z_1 & 0 & 0 & 0
    \\ \cline{2-4} \cline{7-9}
    \multicolumn{9}{c}{} \\[-9pt]
    \multicolumn{1}{c}{} & \multicolumn{3}{c}{$x_3$} &
    &
    \multicolumn{1}{c}{} & \multicolumn{3}{c}{$y_3$}
  \end{tabularx}
  \caption{The projection $\bar\rho$ of $\rho$ onto the reduced game.}
  \label{subfig:ours-reduced-dist}
\end{subfigure}

\caption{The reduced game of Example~\ref{eg:ours}.}
\label{fig:ours-reduced}
\end{figure}

Next, we show in detail that PDCE is not invariant to the duplication of payoff-equivalent strategies in Example~\ref{eg:ours}. To see this, suppose that $w_i$ and $x_i$ are identified as a single strategy $a_i$ for each $i=1,2$, and let $\bar\rho$ be the projection of $\rho$ onto this reduced game. Then $\bar\rho$ assigns probability $1/4$ to each of $(y_1,a_2,x_3)$, $(z_1,a_2,x_3)$, $(a_1,y_2,y_3)$, and $(a_1,z_2,y_3)$. Figure~\ref{fig:ours-reduced} presents the reduced game. 

We claim that $\bar\rho$ is not a PDCE. Suppose, toward a contradiction, that $\bar\rho$ is a PDCE. Then by definition, there exists a sequence of completely mixed conditional strategy profiles $(\sigma^k)$ satisfying the PDCE optimality constraints. For each $k$, let
\[
p^k=\sigma_2^k(y_2\mid a_2),\qquad
q^k=\sigma_2^k(z_2\mid a_2),\qquad
r^k=\sigma_3^k(y_3\mid x_3).
\]
Since $\sigma^k$ is completely mixed, $p^k,q^k,r^k>0$.

Consider player~1. Conditional on being recommended $y_1$, the only recommendation profile in the support of $\bar\rho$ that is consistent with her information is $(y_1,a_2,x_3)$. Thus, under $\sigma^k$, player~2 trembles from $a_2$ to $y_2$ with probability $p^k$ and to $z_2$ with probability $q^k$. Since player~1's payoff from $y_1$ and $z_1$ does not depend on player~3's action in this case, and the common factor $1/4$ cancels from both sides, the optimality constraint comparing $y_1$ with $z_1$ reduces to
\[
p^k-2q^k \ge q^k-2p^k,
\]
or equivalently $p^k\ge q^k$.

Similarly, conditional on being recommended $z_1$, the only recommendation profile in the support of $\bar\rho$ that is consistent with her information is $(z_1,a_2,x_3)$. The optimality constraint comparing $z_1$ with $y_1$ also reduces to
\[
q^k-2p^k \ge p^k-2q^k,
\]
or equivalently $q^k\ge p^k$. Hence $p^k=q^k$. But when $p^k=q^k$, following either $y_1$ or $z_1$ yields payoff
$\frac{1}{4}(p^k-2p^k)=-\frac14p^k<0$.
By contrast, after either recommendation, deviating to $a_1$ yields payoff $\frac{1}{4}r^k(p^k+q^k)=\frac{1}{2}r^kp^k>0$, 
because, with positive probability, player~3 trembles from $x_3$ to $y_3$ and player~2 trembles from $a_2$ to either $y_2$ or $z_2$. This contradicts the optimality constraints. Therefore, we conclude that $\bar\rho$ is not a PDCE.

\end{document}